\documentclass{iwsspa24}
\usepackage{amssymb}
\usepackage[pdftex]{graphicx}
\usepackage{epstopdf}
\usepackage[dvipsnames]{xcolor}
\usepackage{amsmath}
\usepackage{hyperref}
\usepackage[utf8]{inputenc}
\usepackage[bottom]{footmisc}
\usepackage{caption}
\begin{document}

\newcommand{\lm}[1]{{\color{blue} L: #1}}
\newcommand{\ph}[1]{{\color{violet} P: #1}}
\newcommand{\gw}[1]{{\color{teal} G: #1}}

\newcommand{\tentative}[1]{{\color{brown} (#1)}}

\markboth
  {Quantifying Corpus Bias in AMT Systems}
  {Mart\'ak et al.}

\title{Quantifying the Corpus Bias Problem \\in Automatic Music Transcription Systems}

\author{Luk\'a\v{s} Samuel Mart\'ak$^{*}$}{lukas.martak@jku.at}{1}
\author{Patricia Hu$^{*}$}{patricia.hu@jku.at}{1}
\author{Gerhard Widmer}{gerhard.widmer@jku.at}{1}

\affiliation{1}{Institute of Computational Perception \& LIT AI Lab}{Johannes Kepler University, Linz, Austria}

\def\thefootnote{*}\footnotetext{Equal contribution.}
\def\thefootnote{\arabic{footnote}}

\begin{abstract}
Automatic Music Transcription (AMT) is the task of recognizing notes in audio recordings of music. The State-of-the-Art (SotA) benchmarks have been dominated by deep learning systems. Due to the scarcity of high quality data, they are usually trained and evaluated exclusively or predominantly on classical piano music. Unfortunately, that hinders our ability to understand how they generalize to other music.
Previous works have revealed several aspects of memorization and overfitting in these systems. We identify two primary sources of distribution shift: the music, and the sound. Complementing recent results on the sound axis (i.e. acoustics, timbre), we investigate the musical one (i.e. note combinations, dynamics, genre).
We evaluate the performance of several SotA AMT systems on two new experimental test sets which we carefully construct to emulate different levels of musical distribution shift.
Our results reveal a stark performance gap, shedding further light on the Corpus Bias problem, and the extent to which it continues to trouble these systems.

\end{abstract}

\section{Motivation and Goals}

In the past decade, the state of the art in Automatic Music Transcription (AMT) has been repeatedly improved by deploying various Deep Neural Network (DNN) architectures, trained to perform the task in an end-to-end manner~\cite{bock2012polyphonic,sigtia2016endtoend,kelz2016potential,hawthorne2018onsets,kwon2020polyphonic,kong2021highresolution,hawthorne2021sequence,toyama2023automatic}.
Earlier work has shown a tendency of such systems to memorize observed note combinations, hindering their ability to recognize new ones -- called the \emph{entanglement problem}~\cite{kelz2017experimental}.
More recently, the lasting presence of the issue has been observed on a slightly larger scale, denoted as \emph{corpus bias} in~\cite{martak2022balancing}. 
Furthermore, overfitting on sound-related properties of the training data seems to also diminish performance of these systems on out-of-distribution (OOD) data~\cite{hawthorne2019enabling,edwards2024datadriven,hu2024musically}.
We evaluate a set of DNN-based SotA AMT systems, which were trained exclusively on classical music (mostly from  MAESTRO~\cite{hawthorne2019enabling}). %
We name these OaF~\cite{hawthorne2018onsets}, Kong~\cite{kong2021highresolution}, T5~\cite{hawthorne2021sequence}, Toyama~\cite{toyama2023automatic}, and Edwards~\cite{edwards2024datadriven} in the following.
Using a new, highly curated test corpus recorded under uniform \emph{sound} conditions, we contribute new evidence towards the question of AMT performance degradation in the presence of \emph{musical} distribution shift.
To support reproducibility, we make our resources available.\footnote{\url{https://github.com/CPJKU/musical_distribution_shift}}

\section{Methodology}

To produce data for AMT evaluation, we first collect MIDI files and then synthesize them on a real piano, obtaining ground truth note alignments.
In order to eliminate the confounding effects of differences in sound, we construct our corpus by recording automated performances of our MIDI targets on a Yamaha Disklavier grand piano, same as in~\cite{hu2024musically}.
This ensures fixed but realistic timbre, acoustics, and recording conditions of studio quality.
We start by curating two sub-sets: (1) Genre, and (2) Random. 

The Genre set (1) comprises piano performances of pieces from 10 different genres, only one of them being classical music, to investigate the possible effect of this high-level variable on AMT performance.
We source pieces from the ADL Piano MIDI dataset~\cite{ferreira2020computer} as follows: discard pieces containing $>5$ sec of silence or falling outside the range of $2-3$ min duration, and choose 5 pieces per genre at random.

The Random set (2) is synthesized to emulate \emph{extreme} distribution shift, far outside the realm of what a human would call \emph{musical}.
It comprises $24 \times 3$ sequences, each 2 minutes long, maintaining a given polyphony degree $p \in \{1,...,24\}$, and a dynamics range $d \in \{0: [60-68], 1: [32-96], 2: [1-127]\}$.
The $p$ polyphonic streams are filled with note sequences of randomly sampled $\text{pitch} \sim U(21,108)$, $\text{velocity} \sim U(d_{\min},d_{\max})$, and $\text{duration} \sim Beta(\alpha=2,\beta=5)$ between $[0.01 - 5.00]$ seconds.
It differs from the RAND subset of MAPS~\cite{emiya2010multipitch} mainly by randomizing and de-correlating onsets and durations of notes, but also by containing higher polyphony degrees and extra range of dynamics. %

\section{Results and Discussion}

\begin{figure}[h!]
	\centering
	\includegraphics[width=\textwidth]{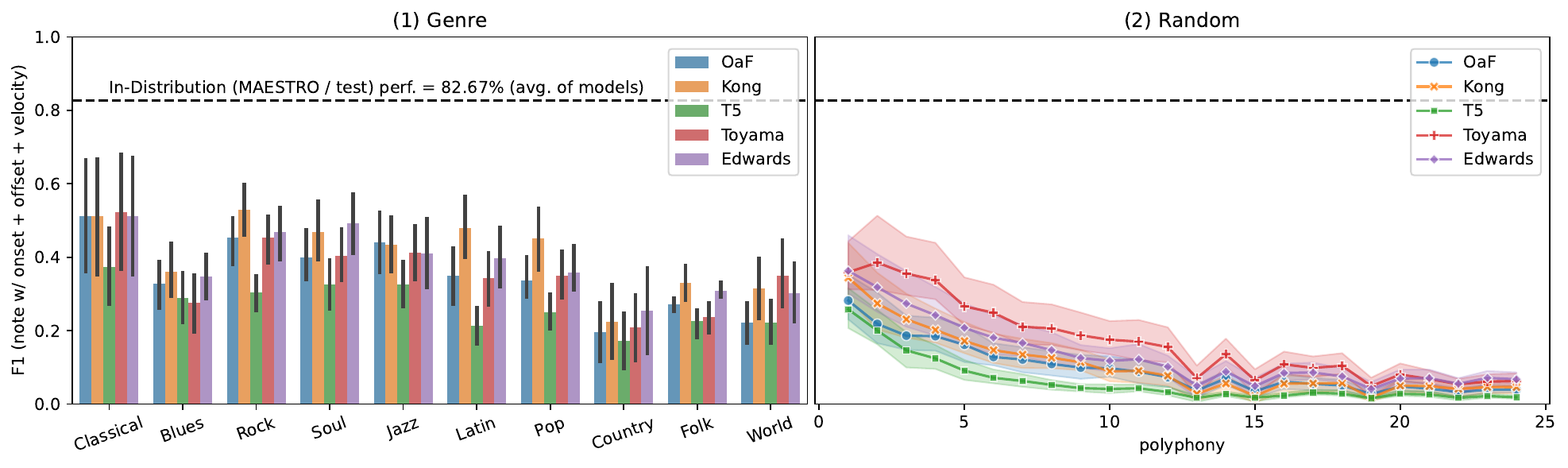}
	\captionsetup{skip=0pt}
	\caption{The Musical OOD performance of SotA systems (Note F1 with Onset, Offset and Velocity).}
	\label{fig:note_on_off_vel}
\end{figure}

We run the systems to transcribe the audio, and report note-level performance by model and data slice in Figure \ref{fig:note_on_off_vel}.
The left plot compares the five models on the different genre test sets (5 pieces per genre; error bars show 50\% confidence intervals).
On the right side, we show the 
performance of each model at different polyphony degrees, averaged over the 3 levels of dynamics, with data bands showing inter-quartile ranges.
The black dashed line gives a rough reference point for average performance across these models on the (in-distribution) MAESTRO test set\textbf{}.
\footnote{The metrics are from the original papers, so the sound distribution for this line is MAESTRO, not ours. We plan to quantify this effect by experiments with (parts of) MAESTRO re-recorded on our piano.}
Most notably, we observe a clear, and partly dramatic, decrease in performance as we go from Classical to other test genres (Fig.~\ref{fig:note_on_off_vel}, left). Performance on random notes (right) tends to be even worse. This demonstrates a strong genre bias effect; in view of this result, performance numbers given in the literature must be interpreted with care.

\newpage

\section*{Acknowledgments}
This work is supported by the European Research Council (ERC) under the EU’s Horizon 2020 research and innovation programme, grant agreement No. 101019375 ("Whither Music?"), and by the Federal State of Upper Austria (LIT AI Lab).

\bibliographystyle{unsrt}
\bibliography{refs}

\end{document}